# A unified approach to truthful scheduling on related machines

Leah Epstein[*]   Asaf Levin[†]   Rob van Stee[‡]

**Abstract**

We present a unified framework for designing deterministic monotone polynomial time approximation schemes (PTAS's) for a wide class of scheduling problems on uniformly related machines. This class includes (among others) minimizing the makespan, maximizing the minimum load, and minimizing the $\ell_p$ norm of the machine loads vector. Previously, this kind of result was only known for the makespan objective. Monotone algorithms have the property that an increase in the speed of a machine cannot decrease the amount of work assigned to it. The *key idea* of our novel method is to show that for goal functions that are sufficiently well-behaved functions of the machine loads, it is possible to compute in polynomial time a highly structured nearly optimal schedule. An interesting aspect of our approach is that, in contrast to all known approximation schemes, we avoid rounding any job sizes or speeds throughout. We can therefore find the *exact* best structured schedule using a dynamic programming. The state space encodes a sufficient amount of information such that no postprocessing is needed, allowing an elegant and relatively simple analysis. The monotonicity is a consequence of the fact that we find the *best* schedule in a specific collection of schedules.

Monotone approximation schemes have an important role in the emerging area of algorithmic mechanism design. In the game-theoretical setting of these scheduling problems there is a social goal, which is one of the objective functions that we study. Each machine is controlled by a selfish single-parameter agent, where its private information is its cost of processing a unit sized job, which is also the inverse of the speed of its machine. Each agent wishes to maximize its own profit, defined as the payment it receives from the mechanism minus its cost for processing all jobs assigned to it, and places a bid which corresponds to its private information. For each one of the problems, we show that we can calculate payments that guarantee truthfulness in an efficient manner. Thus, there exists a dominant strategy where agents report their true speeds, and we show the existence of a truthful mechanism which can be implemented in polynomial time, where the social goal is approximated within a factor of $1 + \varepsilon$ for every $\varepsilon > 0$.

---

[*]Department of Mathematics, University of Haifa, 31905 Haifa, Israel. `lea@math.haifa.ac.il`.
[†]Faculty of Industrial Engineering and Management, The Technion, 32000 Haifa, Israel. `levinas@ie.technion.ac.il`.
[‡]Max Planck Institute for Informatics, Saarbrücken, Germany. `vanstee@mpi-inf.mpg.de`

# 1 Introduction

A major question in algorithmic game theory is how the presence of selfish agents affects the approximability of various classic optimization problems [31]. Specifically, the following research agenda was suggested: *"to what extent is incentive compatible efficient computation fundamentally less powerful than "classic" efficient computation?"* (as formulated in [18]). Of particular interest are scheduling problems, where jobs are assigned for processing to agents, each controlling one machine, and who have some private information regarding their machines [31, 5, 30, 14]. In this paper, we consider the case of single-parameter agents with scheduling problems on uniformly related machines, which was among the first problems considered in the area of algorithmic mechanism design [5]. The private information of an agent is the cost of processing one unit of work, which is also the inverse of the speed of the machine. We provide a negative answer to the question raised in [31] for scheduling problems on uniformly related machines, by designing $(1 + \varepsilon)$-approximation mechanisms for these problems.

Non-preemptive scheduling problems on $m$ uniformly related machines are defined as follows. We let the set of machines be denoted by $M = \{1, 2, \ldots, m\}$. We are given a set of jobs $J = \{1, 2, \ldots, n\}$, where each job $j$ has a positive size $p_j$. The jobs need to be partitioned into $m$ subsets $S_1, \ldots, S_m$, with $S_i$ being the subset of jobs assigned to machine $i$. We let $s_i$ denote the (actual) speed of machine $i$, meaning that the processing of job $j$ takes $\frac{p_j}{s_i}$ time units if $j$ is assigned to machine $i$. For such a solution (also known as a schedule), we let $L_i = (\sum_{j \in S_i} p_j)/s_i$ be the *completion time* or *load* of machine $i$. The *work* of machine $i$ is $W_i = \sum_{j \in S_i} p_j = L_i \cdot s_i$, that is, the total size of the jobs which are assigned to $i$. We consider objective functions which are functions of the machine loads, $L_1, L_2, \ldots, L_m$.

We consider a variety of objective functions (social goals). A well-known objective function is the *makespan*, which is the maximum load. The optimization problem of finding a schedule which minimizes the makespan is a basic one [24, 23, 25, 26, 15]. The problem of finding a schedule which maximizes the minimum load, also known as the *cover*, is the famous *Santa Claus* problem on uniformly related machines (see e.g. [22, 32, 2, 8, 20, 11, 21]). Both these problems are concerned with the optimization of the extremum values of the set of machine loads. We will also consider the optimization problem of minimizing $\sum_{i=1}^{m} f(L_i)$ where $f$ is a well-behaved function. We say that a function $f$ is *well-behaved* if $f$ is a non-negative convex (strictly) monotonically increasing function satisfying the additional property that if $x \leq (1 + \varepsilon)y$ then $f(x) \leq (1 + O(1)\varepsilon)f(y)$. With regard to the problem of minimizing $\sum_{i=1}^{m} f(L_i)$, we assume that there is an oracle such that given a rational number $x$ it computes $f(x)$ exactly in constant time[1]. The most important example of such a function is $f(x) = x^p$ for $p > 1$ in which case the problem is equivalent to minimizing the $\ell_p$ norm of the vector of machines loads. The optimization goal function of minimizing the $\ell_2$ norm (and the goal of minimizing the $\ell_p$ norm for $p > 1$) of the vector of completion times of the machines has been widely studied (see e.g. [17, 13, 7]). The original motivation was minimization of the average latency in storage allocation applications (rather than worst-case latency), and the problem has additional applications in algorithmic game theory [12]. Bansal and Pruhs [10] recently stated: "The standard way to compromise between optimizing for the average and optimizing for the worst case is to optimize the $\ell_p$ norm, generally for something like $p = 2$ or $p = 3$."

The setup of mechanism design for single-parameter agents operating uniformly related machines is as follows. Agents present bids to a mechanism, where the bid $b_i$ of an agent $i$ is the claimed cost per unit of work of its machine (the inverse of its claimed speed). Based on these bids, the mechanism allocates the jobs to the machines and also assigns payments to the agents. We assume that each agent is only interested in maximizing its own profit, which is its payment minus its (actual) cost of processing the jobs allocated to it. A mechanism is called *truthful* if reporting their true costs per unit of work is a dominant strategy for the agents. That is, this strategy maximizes the profit for each agent, regardless of the strategies of the other agents. In the case of single-parameter agents, a well-known necessary and sufficient condition for truthfulness is that the allocation algorithm is *monotone* [5, 4], that is, the allocation algorithm must have the property that if an agent $i$ increases its claimed speed (i.e., decreases its bid) while all other bids are unchanged, the work allocated to $i$ does not decrease. More precisely, in such a case there exist simple payment functions that can be coupled with the (monotone) allocation algorithm to give a truthful mechanism. If the allocation algorithm runs in polynomial time, and the payments can be computed in polynomial time as well, then the resulting truthful mechanism can be implemented in polynomial time. Thus, for single-parameter agents, since the problems are typically strongly NP-hard, the primary goal is to design a monotone (polynomial time) approximation algorithm with the smallest possible approximation ratio, and to show how the corresponding payments can be computed in polynomial time for its outputs.

---
[1] We can loosen this condition by replacing $f$ with a piecewise-linear continuous convex approximation of $f$ (i.e., the approximation is well-behaved as well) without affecting the results. We will assume that $f$ can be computed exactly for simplicity.

An $\mathcal{R}$-approximation algorithm for a minimization problem is a polynomial time algorithm which always finds a feasible solution of cost at most $\mathcal{R}$ times the cost of an optimal solution. An $\mathcal{R}$-approximation algorithm for a maximization problem is a polynomial time algorithm which always finds a feasible solution of value at least $\frac{1}{\mathcal{R}}$ times the value of an optimal solution (we use the convention of approximation ratios greater than 1 for maximization problems). The infimum value of $\mathcal{R}$ for which an algorithm is an $\mathcal{R}$-approximation is called the approximation ratio or the performance guarantee of the algorithm. A polynomial time approximation scheme (PTAS) is a family of approximation algorithms such that the family has a $(1 + \varepsilon)$-approximation algorithm for any $\varepsilon > 0$ (the running time must be polynomial in the input size). If the running time is polynomial in $\frac{1}{\varepsilon}$ as well then the PTAS is in fact an FPTAS (fully polynomial time approximation scheme). On the other hand, if the running time is quasi-polynomial (logarithmic factors of the input size may appear in the exponent), then the approximation scheme (which is not a PTAS) is a quasi-polynomial time approximation scheme (QPTAS). Being strongly NP-hard, the scheduling problems studied here cannot have an FPTAS unless P=NP.

A classic PTAS for these problems generally works by restricting the set of allowable schedules and approximating over this set, where the details depend on the specific algorithm and the objective function considered. Typically, a chief method of restricting allowed schedules is to do grouping and rounding of jobs, where given subsets of jobs are seen as identical, and to treat jobs which are very small compared to the work that a machine should receive as arbitrarily divisible (or sand). A number of difficulties arise when trying to modify such schemes to satisfy the monotonicity requirement (some of which were partially dealt with in the past, see below). It is no longer possible to treat similar jobs as "identical", and their exact sizes must be considered. Jobs which are small for the machine which receives them are much more difficult; such jobs usually do not affect the approximation ratio but which nevertheless need to be assigned very carefully in order to satisfy the monotonicity requirement, since even a very small reduction in the work when the machine increases its speed is not allowed. Moreover, it is not known in advance which job is small on which machine.

Dhangwatnotai et al. [18] used randomization to construct a monotone PTAS for the three main objective functions listed above (makespan, cover, and $\ell_p$ norm), which combined with an appropriate payment function they give, implies a mechanism which is truthful in expectation. That is, given a choice of $\varepsilon > 0$, their algorithm for this value of $\varepsilon$ has an approximation ratio of $1 + \varepsilon$ for *any* realization, but the monotonicity is proved for the expected works of machines. In this weaker notion of truthfulness, the agents are not interested in their actual profits but only in the *expected* ones, that is, the agents are risk-neutral. For example, if an agent earns a profit of $M$ with probability $\frac{1}{M}$ then it sees it as a profit of 1, while a human agent would very much be interested in the value of $M$, and if it is large, it would see it as earning nothing at all (rather than earning 1 in expectation). Their approach of dealing with the difficulties above is that when a machine receives a job of a given rounded size, the actual job is chosen uniformly at random from the set of jobs of this rounded size, so the "sizes" of jobs (the expected sizes) are easier to deal with. For jobs that are small, a fractional assignment is found (and rounded using randomization). They also derived deterministic monotone QPTAS's for minimizing the maximum load and the $\ell_p$ norm of the loads. A fully deterministic (and hence universally truthful) monotone PTAS for minimizing the makespan was given by Christodoulou and Kovács [15]. They assign jobs that have almost the same size (are in the same group) very carefully in a fixed order (sorted by size) to the machines (where machines are given in a fixed order of their speeds). Moreover, they begin by rounding speeds to powers of $1+\varepsilon$, and round the job sizes to powers of $1 + \delta$ for some $\delta \ll \varepsilon$. This ensures that when a speed changes, this change is always relatively large compared to the job classification, so the rounding errors introduced by small jobs are not large compared to the required change in the work. The authors give a long and technical proof to show that it is possible to combine these main ideas and give a deterministic monotone assignment. This approach can be used only for minimizing the makespan, since in the scheme of [15], machines of similar speeds should either receive almost the same work (implied by the makespan), or no small jobs at all, unless no small jobs remain. Informally, the small jobs are pushed to the fastest machines. This approach does not seem to work even for the similar problem of maximizing the cover, but applying the methods of [15] leads to a deterministic monotone $(2 + \varepsilon)$-approximation for this last objective, given by Christodoulou, Kovács, and van Stee [16] (the problem was also studied in [21]).

What can be seen from these previous results is that satisfying the monotonicity requirement would become easier if we could simply avoid the notion of small jobs. Then we could calculate with exact job sizes (and thus exact loads) throughout. An important contribution of this paper is to show that for any given schedule, a highly structured schedule exists, where the ratio of job sizes assigned to a machine is *unbounded* but the jobs types assigned to this machine are restricted in the sense that these jobs are grouped into a sufficiently small number of classes. This overcomes the difficulty that it does not seem to be possible to actually bound the size ratio of jobs assigned to a machine, but still we would like to use dynamic programming *without* introducing a notion of small jobs or inexact calculations. The set of possible outcomes is independent of the possible speeds, which assists in dealing with speed changes, and finally, the work of each machine is very close to its work in the given (original) schedule, which keeps the approximation ratio close to 1. This allows us to deal with *all* of the objective functions mentioned above at once using a dynamic programming formulation implemented by a layered graph, having one layer for each machine. Unlike previous approximation schemes which use such graphs, a path in the graph corresponds to one specific schedule (not to a class of schedules, or a schedule for a set of rounded jobs), and the cost of the path (with respect to a goal function) is precisely the cost of the corresponding schedule and not its approximated value. That is, there is no rounding or imprecise calculation with respect to relatively small jobs (or any other jobs). This makes proving monotonicity much more straightforward, and even simplifies the proof of the approximation ratio, and the presentation of the algorithm, compared to previous (non-monotone) PTAS's. Our construction works in the same way for all inputs and all objectives, and does not require any special cases. Hence we streamline the monotone PTAS for minimizing the makespan [15]. Moreover, we provide the first deterministic monotone PTAS's for maximizing the minimum load and minimizing the $\ell_p$ norm, which are our main contributions.

**Other related work.** For a fixed (constant) number of machines, scheduling problems typically have an FPTAS [27, 9, 19], and even a (deterministic) monotone one for makespan minimization and for maximizing the minimum load [3, 21]. The QPTAS of [18] for minimizing the $\ell_p$ norm is in particular a PTAS for fixed values of $m$. Prior to the monotone FPTAS of Andelman, Azar, and Sorani [3] for makespan minimization, Auletta et al. [6] gave the first deterministic monotone algorithm for this problem (where the number of machines is fixed), with an approximation ratio of $4 + \varepsilon$.

In what follows we discuss the case where the number of machines is part of the input. It was shown by Hochbaum and Shmoys that the makespan minimization problem has a PTAS for identical (equal speed) machines [25] and for uniformly related machines [26]. All optimization problems studied here, including maximizing the minimum load and minimizing the $\ell_p$ norm, are known to have a PTAS for identical machines [32, 1, 2], and for uniformly related machines [8, 20]. As for monotone algorithms for the makespan minimization problem, before the papers [18, 15] mentioned above, Archer and Tardos [5] gave a randomized 3-approximation mechanism for minimizing the makespan which is truthful in expectation only. The ratio was later improved to 2 [4] (and eventually to $1 + \varepsilon$ [18]). A deterministic monotone algorithm of approximation ratio at most 5 was given in [3], and Kovács improved the ratio to 3 and then to 2.8 [28, 29].

**Proof overview.** Our proof consists of two parts. In the first one, we define several properties which a structured schedule should have, and show that every schedule has a similar schedule which has such properties. As stated above, similarity is measured by allowing only a very small change in the work of every machine. For the proof we introduce a notion of a fractional schedule, where some (relatively small) jobs may be split over multiple machines. For any (integral or fractional) schedule, we can define a magnitude vector with a component for every machine. Unlike previous work, where the magnitude of a machine corresponded directly to its work (or the largest job assigned to it), we use the magnitude component of a machine as an upper bound for the size of any job which is assigned to it, but if a component of the magnitude vector is different from the previous one, we require that the value of this component matches (approximately) the work of the corresponding machine. There are several ways to define a magnitude vector for a given schedule. A possible solution to the dynamic programming can be viewed as a process where we create the magnitude vector component by component (for a list of machines sorted by non-decreasing speed); increase the magnitude of the current machine (as opposed to keeping the same magnitude of the previous machine) only if keeping the same magnitude as for the previous machine would result in a violation of the upper bound on the maximum size of any job assigned to the current machine. This novel approach allows additional flexibility in the set of allowed schedules.

For a given integral schedule, where the works of the machines are increasing with the speeds, we show that a fractional schedule exists where the total size of very small jobs which are (partially) assigned to machines with high work is small, and the work on each machine is the same as the work in the integral schedule. We then refine this result by constructing an integral schedule where *no* very small jobs are assigned to machines with high work, the works of the machines are all close to the original works, and an additional technical property holds. However, despite the works being close to the original works, they may no longer be sorted in the resulting schedule (though if the works of two consecutive machines are unsorted, then the difference between their works is very small). Searching for unsorted schedules causes technical difficulties for the algorithm which should find a structured schedule, while a postprocessing step of sorting may harm monotonicity. We therefore do one extra step to create a final integral schedule in which the works are sorted again (but still very close to the original works) and several structural properties hold. We do not use rounding, but jobs are partitioned into mega-classes and mini-classes according to their size, and we apply re-assignment of jobs in every class to comply with the required structure. For a given schedule, some classes of jobs can turn out to be too large for some machines, while they are very small compared to the work of other machines. These jobs are combined into chunks called "alternative jobs". Since this process can be applied in particular for an optimal schedule (for each one of the studied problems), there exists a schedule where works are very close to the works in an optimal schedule, and the structured schedule has an objective value which is close to optimum.

Once we show the existence of such a schedule, we can turn to the design of an algorithm which finds it. We use a dynamic programming formulation which is based on the structural properties. By the structural properties and the existence of a magnitude vector, it is only necessary to have a small number of components of this vector in the state space. A preprocessing step is performed, where all possible types of alternative jobs are created. While a job will belong to a number of sets of alternative jobs, every solution will use it at most once as a part of an alternative job (or possibly it will simply be assigned as a job). Thus, we find an optimal solution out of a given class using a polynomial time algorithm, and this optimal schedule is then guaranteed to be close to an overall optimal schedule, as well as being monotone.

## 2 Preliminaries

For our results, we let $\varepsilon$ be a small constant such that $0 < \varepsilon \leq \frac{1}{32}$ and $\frac{1}{\varepsilon}$ is an integer power of 2 denoted by $r \geq 5$ (i.e., $\varepsilon = \frac{1}{2^r}$). Throughout the paper, for a solution $\mathcal{A}$ we denote by $\mathcal{A}$ both the solution and the value of the objective function for this solution. Without loss of generality, we assume that $0 < p_1 \leq p_2 \leq \cdots \leq p_n$.

An integral schedule is a function $S : J \to M$. We let $W_i^S = \sum_{j \in J : S(j) = i} p_j$ (this is the work of machine $i$ in the integral schedule $S$). A fractional schedule is a function $X : J \times M \to [0, 1]$. The value $X(j, i)$ is the fraction of job $j$ assigned to machine $i$, and the following condition (that every job is assigned completely) must be satisfied:

**(F1)** For every $j \in J$, $\sum_{i \in M} X(j, i) = 1$.

Let $W_i^X = \sum_{j \in J} p_j \cdot X(j, i)$ be the total fractional size of jobs of machine $i$, and let $\tilde{W}_i^X = 2^{\alpha_i^X}$, where $\alpha_i^X = \lceil \log_2 W_i^X \rceil$, be its rounded value (if $W_i^X = 0$ then $\alpha_i^X = -\infty$ and $\tilde{W}_i^X = 0$). We call $W_i^X$ the work of machine $i$ in $X$ (as for integral schedules) and $\tilde{W}_i^X$ is the rounded work (also for integral schedules). A fractional schedule is *valid* if it satisfies condition (F2):

**(F2)** There is a partition $J = J_\mathbb{Z}(X) \cup J_\mathbb{R}(X)$ ($J_\mathbb{Z}(X) \cap J_\mathbb{R}(X) = \emptyset$), such that if $j \in J_\mathbb{Z}(X)$ then there is a unique value $i \in M$ such that $X(j, i) > 0$ (and therefore $X(j, i) = 1$), and if $j \in J_\mathbb{R}(X)$ and $X(j, i) > 0$ then $p_j \leq \varepsilon \tilde{W}_i^X$.

Note that the partition in (F2) is not necessarily uniquely defined. Every integral schedule $S$ induces a valid fractional schedule $X$ with the same jobs assigned to every machine as follows: let $X(j, i) = 1$ if $S(j) = i$, else $X(j, i) = 0$. Furthermore, we let $J_\mathbb{R}(X) = \{j \in J : p_j \leq \varepsilon \tilde{W}_{S(j)}^S\}$ and $J_\mathbb{Z}(X) = J \setminus J_\mathbb{R}(X)$. Note that $\tilde{W}_i^S = \tilde{W}_i^X$ for $i = 1, \ldots, m$. $X$ is called the (valid) fractional schedule induced by $S$. On the other hand, every valid fractional schedule $X$ for which $X(j, i) \in \{0, 1\}$ for all $j \in J, i \in M$ induces an integral schedule $S$ with the same works by setting $S(j) = i$ for the value of $i$ for which $X(j, i) = 1$ (this value of $i$ is unique due to (F1)). $S$ is called the integral schedule induced by $X$. In what follows we use the term *schedule* for an integral schedule. We let $L_i^S = \frac{W_i^S}{s_i}$ be the load of machine $i$ in the schedule $S$. The first part of Claim 1 follows from an observation in [20], and it is easy to show the second part (see Appendix A.1). For all cases, we conclude that if machines are sorted by non-decreasing speed, it is sufficient to consider optimal schedules where the works are non-decreasing (as a function of the indices).

**Claim 1** *Assume that $s_1 \leq s_2 \leq \cdots \leq s_m$. There exists an optimal schedule $S$ for the problem of minimizing $\sum_{i=1}^{m} f(L_i)$ where $f$ is a well-behaved function, which satisfies $W_1^S \leq W_2^S \leq \cdots \leq W_m^S$. There exists an optimal schedule $S_1$ for the makespan minimization problem which satisfies $W_1^{S_1} \leq W_2^{S_1} \leq \cdots \leq W_m^{S_1}$. There exists an optimal schedule $S_2$ for the machine covering problem which satisfies $W_1^{S_2} \leq W_2^{S_2} \leq \cdots \leq W_m^{S_2}$.*

## 3 The existence of near-optimal highly structured solutions

We define a partition of $J$ into mega-classes. For $k \in \mathbb{Z}$, let $\mathcal{I}_k = (2^k, 2^{k+1}]$, and let mega-class $k$ be $\{j \in J : p_j \in \mathcal{I}_k\}$. We say that an integer $k$ *dominates* the integer $k'$ if $k > k' + r$. Mega-class $k$ *dominates* mega-class $k'$ if $k$ dominates $k'$. If $j, j'$ belong to mega-classes $k, k'$, respectively, such that mega-class $k$ dominates mega-class $k'$, then $p_{j'} < \varepsilon p_j$. This holds because $p_j > 2^k \geq 2^{k'+r+1} = \frac{1}{\varepsilon} \cdot 2^{k'+1} \geq \frac{1}{\varepsilon} \cdot p_{j'}$, since $k' + 1 \leq k - r$ and $\varepsilon = 2^{-r}$. We refine this partition and consider the partition of $J$ into mini-classes as follows. Denote by $K \subseteq \mathbb{Z}$ the set of indices of non-empty mega-classes (clearly $|K| \leq n$). Let $\lambda = \lceil \log_{1+\varepsilon} 2 \rceil$. For $k \in K$ and $0 \leq \ell \leq \lambda - 1$, let $I_{k,\ell} = (2^k \cdot (1+\varepsilon)^\ell, 2^k \cdot (1+\varepsilon)^{\ell+1}]$. The mini-class $(k, \ell)$ is the set of jobs of mega-class $k$ whose size is in $I_{k,\ell}$. Note that $(1+\varepsilon)^{\lceil \log_{1+\varepsilon} 2 \rceil} \geq (1+\varepsilon)^{\log_{1+\varepsilon} 2} = 2$ and thus the partition of $J$ into the mini-classes is a refined partition of the partition into the mega-classes.

Given a set of consecutive mega-classes $k_1, \ldots, k_2$ where $k_2 \geq k_1$, with the job set $\hat{J}$ consisting of all jobs of $J$ with size in the interval $(2^{k_1}, 2^{k_2+1}]$, and letting $\varrho = 2^{k_2}$, we create an alternative set of jobs that will possibly replace $\hat{J}$. These alternative jobs have size in the interval $(\varrho, 2\varrho]$ (except perhaps for one alternative job that may be smaller). To create these alternative jobs we partition $\hat{J}$ into subsets each of which has total size at most $2\varrho$ such that no two subsets can be united keeping this condition. A set of subsets satisfying this condition has at most one subset whose total size is at most $\varrho$. We create these subsets by picking in each step a maximal prefix of the jobs in $\hat{J}$ (where $\hat{J}$ is sorted according to the indices of the jobs, i.e., by non-decreasing size) with total size at most $2\varrho$ and remove the selected jobs from $\hat{J}$. This algorithm is equivalent to applying the bin packing algorithm Next-Fit Increasing (NFI) using "bins" of size $2\varrho$; once a subset of total size at most $\varrho$ is picked, all further subsets (if any exist) have total sizes above $\varrho$. The algorithm sometimes decides to replace $\hat{J}$ with the alternative jobs, and in this case we partition these alternative jobs into separate mini-classes which we call alternative mini-classes. The alternative mini-class $(k, \ell)$ contains all the alternative jobs with size in $I_{k,\ell}$, resulting in at most $\lambda + 1$ alternative mini-classes. If the algorithm decides to replace $\hat{J}$ with alternative jobs, then in the output of the algorithm each alternative job is replaced with the original jobs which were combined to form it, and this is done just before returning the output (the work of each machine is not affected by this change). Since there are at most $n$ non-empty mega-classes, there are $O(n^2)$ different sets $\hat{J}$ that possibly the algorithm replaces with alternative jobs. Thus creating all the sets of alternative jobs takes $O(n^3)$. Note that one job can be contained in multiple alternative jobs, but at most one of these alternative jobs will be used.

**Definition 2** *An integral schedule* respects *the alternative jobs of mega-classes $k_1, \ldots, k_2$, where $k_2 \geq k_1$,*

*if every pair of jobs $j, j'$ with size in the interval $(2^{k_1}, 2^{k_2+1}]$ which are within a common subset (that is, should be combined into one alternative job with possibly other jobs), are scheduled on a common machine.*

The motivation for this definition is that these jobs can be easily replaced by the alternative job to which they belong without affecting the works of the machines.

**Definition 3** *A vector $\bar{a} = (a_0, a_1, \ldots, a_m)$ (of length $m+1$) whose components belong to $\mathbb{Z} \cup \{-\infty\}$ is called a* magnitude vector *if $a_0 = -\infty$, for $i = 0, 1, \ldots, m-1$, $a_i \leq a_{i+1}$ and if $a_i \neq a_{i+1}$ then $a_{i+1}$ dominates $a_i$ (i.e., $a_{i+1} \geq a_i + r + 1$).*

We now define the *signature vector* $\bar{b}$ of a magnitude vector $\bar{a}$. The number of components in $\bar{b}$ is the number of distinct values among the components of $\bar{a}$ excluding $a_0$, denoted by $\tau(\bar{a})$. Each component $t = 1, 2, \ldots, \tau(\bar{a})$ of $\bar{b}$ is a pair $b_t = (\xi_t, \nu_t)$ such that $\xi_1 = 1$, and for $1 \leq t \leq \tau(\bar{a})$ and $\xi_t \leq i \leq \xi_{t+1} - 1$ (where $\xi_{\tau(\bar{a})+1} = m + 1$) we have $a_i = \nu_t$. That is, the value $\xi_t$ is always the first machine which has a larger component of $\bar{a}$ than the previous machine and this component is $\nu_t$. For every $t = 1, 2, \ldots, \tau(\bar{a}) - 1$, we let $J^t(\bar{a}) = \{j \in J : 2^{\nu_t + r + 1} < p_j \leq 2^{\nu_{t+1} - r}\}$.

**Observation 4** *For every job $j$ and every magnitude vector $\bar{a}$ with its signature vector $\bar{b}$, there are at most two values of $t \in \{1, \ldots, \tau(\bar{a})\}$ for which $p_j \in (2^{\nu_t - r}, 2^{\nu_t + r + 1}]$, and if there exists at least one such value of $t$, then $j \notin \cup_\theta J^\theta(\bar{a})$.*

**Proof.** By the definitions above, for every $j$ there are at most two values of $t$ for which $p_j \in (2^{\nu_t - r}, 2^{\nu_t + r + 1}]$ (since $\nu_{\theta+1} \geq \nu_\theta + r + 1$ for every $\theta$). Moreover, if $p_j \in (2^{\nu_t - r}, 2^{\nu_t + r + 1}]$, then for every $\theta < t$, we have $j \notin J^\theta(\bar{a})$ because $p_j > 2^{\nu_t - r} \geq 2^{\nu_{\theta+1} - r}$, and thus $j$ is too large to be in $J^\theta(\bar{a})$. If $\theta \geq t$, then $p_j \leq 2^{\nu_t + r + 1} \leq 2^{\nu_\theta + r + 1}$, and thus $j$ is too small to be in $J^\theta(\bar{a})$. ∎

**Definition 5** *A valid fractional schedule $X$ is* consistent *with a magnitude vector $\bar{a}$ if 1) for every job $j$ and machine $i$, if $X(j, i) > 0$ then $p_j \leq 2^{a_i + r + 1}$, that is, machine $i$ does not contain parts of jobs of a mega-class higher than $a_i + r$, and 2) if $a_i \neq a_{i-1}$ (for $i \in M$) then $a_i = \alpha_i^X (= \lceil \log_2 W_i^X \rceil)$.*

**Observation 6** *If a valid fractional schedule $X$ is consistent with a magnitude vector $\bar{a}$ and $W_1^X \leq W_2^X \leq \cdots \leq W_m^X$, then for every $i \in M$, we have $a_i \leq \alpha_i^X$.*

**Definition 7** *A pair $(X, \bar{a})$, where $X$ is a valid fractional schedule, and $\bar{a}$ is a magnitude vector such that $X$ is consistent with $\bar{a}$ is called* favorable *if for $t = 1, 2, \ldots, \tau(\bar{a}) - 3$, we have*

$$\sum_{i=\xi_{t+3}}^{m} \sum_{j : p_j \leq 2^{\nu_t - r}} p_j \cdot X(j, i) \leq 2^{\nu_{t+1} + r + 1} .$$

This condition ensures in particular that the total size of parts of jobs whose mega-class is dominated by mega-class $\nu_t$, assigned to a machine of index at least $\xi_{t+3}$, is relatively small compared to the work of that machine. This holds since $2^{\nu_{t+1} + r + 1} \leq 2^{\nu_{t+2}} < 2^{\nu_{t+3} - r} = \varepsilon \cdot 2^{\nu_{t+3}} = \varepsilon \cdot \tilde{W}_{\xi_{t+3}}^X$.

We define several processes in which a valid fractional schedule is modified into a different valid fractional schedule. These processes are defined algorithmically but they are not a part of the final algorithm, but only of the proof that a highly structured integral schedule must exist.

**FNFI.** For a subset of jobs $J' \subseteq J$ and a set of bounds $U_1, \ldots, U_m$ (for the $m$ machines) such that $\sum_{j \in J'} p_j = \sum_{i=1}^m U_i$, the Fractional Next-Fit Increasing (FNFI) algorithm creates a fractional allocation of these jobs in the following way. Let $i = 1$ be the first active machine, and for every $j \in J'$ let $q_j = p_j$. In every step, FNFI picks the minimum index job $j \in J'$. It allocates $\beta = \min\{q_j, U_i\}$ processing time of this job to machine $i$. It decreases both $U_i$ and $q_j$ by $\beta$. If $U_i = 0$, then it increases $i$ by 1, and if $q_j = 0$, then it removes $j$ from $J'$. FNFI repeats this step until $i = m+1$ (and $J' = \emptyset$ must hold, these two events happen simultaneously since $\sum_{j \in J'} p_j = \sum_{i=1}^m U_i$). FNFI is sometimes used to reassign a subset of jobs in a valid fractional schedule, so that the total sizes of jobs of this subset assigned to each machine is unchanged (i.e., the bounds $U_i$ are given by assignment of the jobs of the subset in the original valid fractional schedule). This is done only in situations where it is ensured that the resulting fractional schedule is valid.

**Definition 8** *A valid fractional schedule $X$ is* compatible with FNFI *if running* FNFI *on the input job set $J_\mathbb{R}(X)$ with the set of bounds $U_1, \ldots, U_m$ such that $U_i = \sum_{j \in J_\mathbb{R}(X)} p_j X(j, i)$ allocates exactly $p_j \cdot X(j, i)$ time units of job $j$ to machine $i$ for every $j \in J_\mathbb{R}(X)$ and all $i \in M$, that is, keeps the valid fractional schedule unchanged.*

**Round-FNFI.** On several occasions, given a valid fractional schedule $X$, which is compatible with FNFI, we will apply the following rounding procedure, called Round-FNFI. Assign each job $j \in J_\mathbb{R}(X)$ completely to the minimum index $i$ such that $X(j, i) > 0$. Since in the assignment process of FNFI each machine receives at most two jobs which are not completely assigned to it, the one of smallest index and the one of largest index, the resulting fractional schedule induces an integral schedule $S$ in which each machine may have additional parts of at most one job (the one of the largest index assigned to this machine by FNFI), and may have less parts of at most one job (the one of the smallest index assigned to this machine by FNFI). Since by condition (F2) each fractional job $j \in J_\mathbb{R}(X)$ on machine $i$ (that is, every $j \in J_\mathbb{R}(X)$ such that $X(j, i) > 0$) has size $p_j \leq \varepsilon \tilde{W}_i^X \leq 2\varepsilon W_i^X$, we conclude that for every $i \in M$ we have $(1 - 2\varepsilon) W_i^X \leq W_i^S \leq (1 + 2\varepsilon) W_i^X$. We say that the integral schedule $S$ is created by applying Round-FNFI on $X$.

**Lemma 9** *Given a schedule $S : J \to M$ such that $W_1^S \leq W_2^S \leq \cdots \leq W_m^S$, there exists a favorable pair $(X, \bar{a})$ where $W_i^X = W_i^S$ for $i = 1, 2, \ldots, m$, and $X$ is compatible with FNFI.*

**Proof.** First, as described in Section 2, $S$ induces a valid fractional schedule, here denoted by $X_S$, with the same sequence of works. For $i = 1, 2, \ldots, m$, let $q_i = \alpha_i^S$. Since $W_1^S \leq W_2^S \leq \cdots \leq W_m^S$, we have $q_i \leq q_{i+1}$ for all $i = 1, 2, \ldots, m-1$. We define a magnitude vector $\bar{a}^S = (a_0^S, a_1^S, \ldots, a_m^S)$ as follows. We let $a_0^S = -\infty$, for $i = 1, 2, \ldots, m$, if $q_i \leq a_{i-1}^S + r$, then $a_i^S = a_{i-1}^S$, and otherwise $a_i^S = q_i$. The valid fractional schedule $X_S$ is consistent with $\bar{a}^S$ since for every $i \in M$ either $a_i^S = q_i$ or $q_i \leq a_{i-1}^S + r = a_i^S + r$. In both cases, the size of any job assigned (completely) to machine $i$ cannot exceed $W_i^X \leq \tilde{W}_i^X = 2^{q_i} \leq 2^{a_i^S + r}$.

We next consider the nonempty set of pairs $(X', \bar{a})$ such that $X'$ is consistent with $\bar{a}$, and such that for $i = 1, 2, \ldots, m$, $W_i^{X'} = W_i^S$ and $a_i \leq \alpha_i^{X'}$ (the set is indeed nonempty by the existence of $(X_S, \bar{a}^S)$). Among all the possible choices for $X'$ and $\bar{a}$, we consider one such that the vector $\bar{a}$ has a signature vector with the smallest number of components, and (as a secondary objective, i.e., among such solutions which minimize the number of components in the signature vector) $|J_\mathbb{R}(X')|$ is maximized. Based on $X'$ we will define $X$ (by applying FNFI on $J_\mathbb{R}(X')$), and $X$ will be shown to be a valid fractional schedule satisfying the lemma.

We modify $X'$ by reassigning the jobs of $J_\mathbb{R}(X')$ using FNFI with the set of bounds $U_1, \ldots, U_m$ such that $U_i = \sum_{j \in J_\mathbb{R}(X')} p_j X'(j,i)$. We denote the resulting fractional schedule which is compatible with FNFI by $X$. We argue that $X$ satisfies (F2). We define $J_\mathbb{R}(X) = J_\mathbb{R}(X')$ and show that if $j \in J_\mathbb{R}(X)$ and $i \in M$ satisfy that $X(j,i) > 0$, then $p_j \leq \varepsilon \tilde{W}_i^X$. Since the works of the machines are sorted in a non-decreasing order, it suffices to show that for $j \in J_\mathbb{R}(X)$ and $i$ such that $X(j,i) > 0$, there exists $j' \geq j$, $j' \in J_\mathbb{R}(X')$ and $i' \leq i$ such that $X'(j', i') > 0$, since in such a case $p_j \leq p_{j'} \leq \varepsilon \tilde{W}_{i'}^{X'} \leq \varepsilon \tilde{W}_i^X$. Assume by contradiction that this claim does not hold for $j$ and $i$. Then, since FNFI assigns job $j$ (possibly partially) to machine $i$, $\sum_{j' \in J_\mathbb{R}(X'):j'<j} p_{j'} < \sum_{\gamma=1}^{i} U_\gamma$, however $\sum_{j' \in J_\mathbb{R}(X'):j'<j} p_{j'} \geq \sum_{\gamma=1}^{i} U_\gamma$ since no other jobs of $J_\mathbb{R}(X')$ are assigned by $X'$ to the first $i$ machines. Therefore $X$ is indeed a valid fractional schedule.

We claim that $X$ is consistent with $\bar{a}$. It suffices to prove that in every prefix of machines $1, 2, \ldots, i$, the maximum size of a job $j$ such that $X(j, \gamma) > 0$ for some $1 \leq \gamma \leq i$ does not increase when we replace $X'$ by $X$. Let $j$ be a job of maximum size which is assigned in $X$ (possibly fractionally) to a machine $\gamma \in \{1, 2, \ldots, i\}$. If $j \in J_\mathbb{Z}(X) = J_\mathbb{Z}(X')$ then $X'(j, \gamma) = X(j, \gamma) = 1$, and the claim holds. Otherwise, $j \in J_\mathbb{R}(X)$. There exists $j' \in J_\mathbb{R}(X')$ and $i' \leq \gamma$ such that $X'(j', i') > 0$ and $j' \geq j$ as we showed above, and the claim holds as well.

Last, we prove that $(X, \bar{a})$ is a favorable pair. Let $t$ be such that $1 \leq t \leq \tau(\bar{a}) - 3$. Let $j \in J$ be such that there is $i \in [\xi_{t+3}, m]$ with $X(j,i) > 0$ and $p_j \leq 2^{\nu_t - r}$. If there is no such job, then we are done. We have $j \in J_\mathbb{R}(X) = J_\mathbb{R}(X')$ because $\tilde{W}_i^X \geq 2^{a_i} \geq 2^{\nu_{t+3}} > 2^{\nu_t} \cdot 2^{3r} = \frac{1}{\varepsilon^3} \cdot 2^{\nu_t} \geq \frac{1}{\varepsilon^3} p_j$ where the first inequality holds by Observation 6, so if $j \notin J_\mathbb{R}(X)$ then $X(j,i) = 1$ and we can add $j$ to $J_\mathbb{R}(X)$, contradicting our choice of $X'$. Consider the machines $A_{t+1} = \{\xi_{t+1}, \ldots, \xi_{t+2} - 1\}$. If all jobs assigned (possibly fractionally) by $X$ to these machines have size of at most $2^{\nu_t + r + 1}$, then we can redefine $a_{i'}$ for $i' \in A_{t+1}$ to be $\nu_t$, contradicting the minimality of length of the signature vector of $\bar{a}$. Consider a job $j'$ such that there is $i' \in A_{t+1}$ for which $X(j', i') > 0$ and $p_{j'} > 2^{\nu_t + r + 1}$. By the existence of $j \in J_\mathbb{R}(X)$ with size at most $2^{\nu_t - r}$, such that a part of it is allocated to a machine of higher index, we conclude that $j' \in J_\mathbb{Z}(X)$ since $X$ is compatible with FNFI. We also have $p_{j'} \leq 2^{\nu_{t+1} + r + 1} \leq 2^{\nu_{t+3} - r - 1} < \varepsilon \cdot \tilde{W}_{\xi_{t+3}}^X$. If $\sum_{\gamma = \xi_{t+3}}^{m} \sum_{j'': p_{j''} \leq 2^{\nu_t - r}} p_{j''} \cdot X(j'', \gamma) > 2^{\nu_{t+1} + r + 1}$, then $\sum_{\gamma = \xi_{t+3}}^{m} \sum_{j'': p_{j''} \leq 2^{\nu_t - r}} p_{j''} \cdot X(j'', \gamma) > p_{j'}$. In this case, we add $j'$ to $J_\mathbb{R}(X)$, and modify $X$ as follows. We consider a replacement of the position of $j'$ with the position of a set of fractions of jobs (where each such job has size at most $2^{\nu_t - r} = \varepsilon 2^{\nu_t} \leq \varepsilon^2 2^{\nu_{t+1} - 1} \leq \varepsilon^2 \frac{\tilde{W}_\gamma^X}{2} < \varepsilon^2 W_\gamma^X$ for every $\gamma \in A_{t+1}$, and belongs to $J_\mathbb{R}(X')$) of total size $p_{j'}$ which were previously assigned to machines with index at least $\xi_{t+3}$. The resulting schedule indeed satisfies (F2) since the jobs which take the place of $j'$ are smaller than $\varepsilon^2 W_\gamma^X$ for every $\gamma \in A_{t+1}$ while $p_{j'} < \varepsilon \cdot \tilde{W}_{\xi_{t+3}}^X$. Thus, the resulting valid fractional schedule is consistent with $\bar{a}$, contradicting our choice of $X'$ since $|J_\mathbb{R}(X')|$ is not maximal among valid fractional schedules consistent with $\bar{a}$ (and having the required properties). ∎

**Definition 10** *A schedule $S$ is* almost consistent *with a magnitude vector $\bar{a}$ if for every $i = 1, 2, \ldots, m$, the set of parts of jobs assigned to machine $i$ does not contain any part of a job of a mega-class higher than $a_i + r$, and if $a_i \neq a_{i-1}$ (for $i \in M$) then $|a_i - \alpha_i^S| \leq 1$.*

**Definition 11** *A schedule $S : J \to M$ is* good *if the following properties hold.*

1. *There exists a magnitude vector $\bar{a}$ such that $S$ is almost consistent with $\bar{a}$, and furthermore for every $t = 1, 2, \ldots, \tau(\bar{a}) - 4$ there is no $j$ and $i \geq \xi_{t+4}$ such that $p_j \leq 2^{\nu_t - r}$ and $S(j) = i$.*

2. *For every $t = 1, 2, \ldots, \tau(\bar{a}) - 1$ if $J^t(\bar{a}) = \{j \in J : 2^{\nu_t + r + 1} < p_j \leq 2^{\nu_{t+1} - r}\} \neq \emptyset$, then $X$ respects the alternative jobs of mega-classes $\nu_t + r + 1, \ldots, \nu_{t+1} - r - 1$.*

**Lemma 12** *Given a schedule* $S : J \to M$ *such that* $W_1^S \leq W_2^S \leq \cdots \leq W_m^S$, *there exists a good schedule* $S' : J \to M$ *such that for* $i = 1, 2, \ldots, m$, *we have*

$$(1 - 12\varepsilon) \cdot W_i^S \leq W_i^{S'} \leq (1 + 12\varepsilon) \cdot W_i^S. \tag{1}$$

**Proof.** By Lemma 9, there exists a favorable pair $(X, \bar{a})$ where $W_i^X = W_i^S$ for $i = 1, 2, \ldots, m$, and $X$ is compatible with FNFI. First, for every $t = 4, 5, \ldots, \tau(\bar{a})$, we reschedule all parts of jobs $j$ such that $p_j \leq 2^{\nu_t - 3 - r}$ and for which there exists $i > \xi_t$ such that $X(j, i) > 0$ by moving them to machine $\xi_t$. We denote by $\tilde{X}$ the resulting fractional schedule. We next bound the value of $W_i^{\tilde{X}}$ in terms of $W_i^X$ for every $i \in M$. The work of $i$ may increase (if $i = \xi_t$ for some $t = 4, 5, \ldots, \tau(\bar{a})$). Since $(X, \bar{a})$ is a favorable pair, the amount of this increase is at most $\varepsilon \cdot \tilde{W}_i^X < 2\varepsilon W_i^X$, since $2^{\nu_t - 2 + r + 1} < 2^{\nu_t - r} = \varepsilon \tilde{W}_i^X$. Next, we bound the total size of parts of jobs removed from machine $i$ (for $2 \leq i \leq m$). Let $t'$ be the maximum index such that $\xi_{t'} < i$ (which must exist since $\xi_1 = 1$). Then, for every $t = 4, 5, \ldots, t'$, we may have removed a total size of at most $2^{\nu_t - 2 + r + 1} \leq \frac{\varepsilon}{2} \cdot 2^{\nu_t}$ from machine $i$ (and move these parts of jobs to machine $\xi_t$). Thus $W_i^X - W_i^{\tilde{X}} \leq \frac{\varepsilon}{2} \cdot \sum_{t=4}^{t'} 2^{\nu_t} \leq \varepsilon \cdot 2^{\nu_{t'}} \leq 2\varepsilon \cdot W_i^X$. We conclude that for every $i$, we have $(1 - 2\varepsilon) W_i^X \leq W_i^{\tilde{X}} \leq (1 + 2\varepsilon) W_i^X$.

Let $J_\mathbb{R}(\tilde{X}) = J_\mathbb{R}(X)$. We observe that $\tilde{X}$ is a valid fractional schedule which is compatible with FNFI (similarly to the bounds on such jobs in Lemma 9, it can be shown that if a job moved to machine $i$, then its size is below $\varepsilon W_i^{\tilde{X}}$, since $1 - 2\varepsilon > \varepsilon$). We now apply Round-FNFI on $\tilde{X}$ to create an integral schedule $\tilde{S}$. Every $j \in J_\mathbb{R}(\tilde{X})$ such that $\tilde{X}(j, i) > 0$ has size $p_j \leq \varepsilon \tilde{W}_i^X \leq 2\varepsilon W_i^X$, so for every $i \in M$ we have

$$(1 - 4\varepsilon) W_i^X \leq W_i^{\tilde{S}} \leq (1 + 4\varepsilon) W_i^X . \tag{2}$$

The maximum size of a job in a prefix of machines in $\tilde{S}$ is the same as in $\tilde{X}$, and a job moved from its position in $X$ to a new position on machine $\xi_t$ in $\tilde{X}$ has size at most $2^{\nu_t - 3} < \varepsilon 2^{\nu_t} = \varepsilon 2^{a_{\xi_t}}$.

Consider the set of jobs $J^t(\bar{a})$. Since $X$ is consistent with $\bar{a}$, for every $j \in J^t(\bar{a})$ and $i < \xi_{t+1}$, we have $X(j, i) = 0$, and since the maximum size of a job in a prefix of machines did not change, $\tilde{S}(j) > i$. Since $\tilde{W}_{\xi_{t+1}}^X = 2^{\nu_{t+1}}$, we have for all $j \in J^t(\bar{a})$ and $i \geq \xi_{t+1}$ that $p_j \leq 2^{\nu_{t+1} - r} = \varepsilon \cdot \tilde{W}_{\xi_{t+1}}^X \leq \varepsilon \cdot \tilde{W}_i^X$. We remove the jobs in $J^t(\bar{a})$ from their positions in $\tilde{S}$, and we will schedule the alternative jobs instead (which gives a schedule of the original jobs which respects the alternative jobs of mega-classes $\nu_t + r + 1, \ldots, \nu_{t+1} - r - 1$). For every $i \in M$ we let $U_i$ be the total size of jobs in $J^t(\bar{a})$ which are assigned to machine $i$ by $\tilde{S}$. The set of machines $i$ for which $U_i \neq 0$ is contained in the interval $[\xi_{t+1}, \xi_{t+4}]$ where if $t + 4 > \tau(\bar{a})$, then we let $\xi_{t+4} = m$. We apply FNFI to fractionally schedule the alternative jobs, followed by Round-FNFI. This is done for every value of $t$ for which $J^t(\bar{a}) \neq \emptyset$ sequentially. We denote by $S'$ the resulting integral solution. Let $i \in M$. There are at most four values of $t$ for which $i$ participated in the process of the rescheduling of $J^t(\bar{a})$. As a result of applying Round-FNFI for the alternative jobs for all $t$, every machine $i$ can have at most four additional parts of jobs and less parts of at most four jobs, all of which have size of at most $\varepsilon \tilde{W}_i^X \leq 2\varepsilon W_i^X$. Thus, $W_i^{\tilde{S}} - 8\varepsilon W_i^X \leq W_i^{S'} \leq W_i^{\tilde{S}} + 8\varepsilon W_i^X$. Using (2), we get (1).

The integral schedule $S'$ is almost consistent with the magnitude vector $\bar{a}$. To see this claim, first observe that no job is too large: if the maximum size of a job on machine $i$ in $S'$ is not the same as in $\tilde{S}$, this maximum size job $j \in J^t(\bar{a})$ is moved from its position in $\tilde{S}$ to a new position on machine $i$, and therefore $p_j \leq 2^{\nu_{t+1} - r} = \varepsilon 2^{\nu_{t+1}}$, and $a_i \geq a_{\xi_{t+1}} = \nu_{t+1}$. The claim holds because for every $i$, we have $|\alpha_i^X - \alpha_i^{S'}| \leq 1$ since $1 + 12\varepsilon < 2$ and $\frac{1}{1 - 12\varepsilon} < 2$. ∎

**Definition 13** *A schedule $S$ is* quasi-consistent *with a magnitude vector $\bar{a}$ if for every $i = 1, 2, \ldots, m$ such that $\xi_t \leq i < \xi_{t+1}$, the set of jobs assigned to machine $i$ does not contain any job of a mega-class higher than $\nu_{t+1} + r$, and if $a_i \neq a_{i-1}$ (for $i \in M$) then $|a_i - \alpha_i^S| \leq 1$.*

**Definition 14** *A schedule $S : J \to M$ is* structured *if the following properties hold.*

1. *There exists a magnitude vector $\bar{a}$ such that $S$ is quasi-consistent with $\bar{a}$, and furthermore for every $t = 1, 2, \ldots, \tau(\bar{a}) - 5$ there is no $j$ and $i \geq \xi_{t+5}$ such that $p_j \leq 2^{\nu_t - r}$ and $S(j) = i$.*

2. *For every $t = 1, 2, \ldots, \tau(\bar{a}) - 1$, if $J^t(\bar{a}) \neq \emptyset$, then $S$ respects the alternative jobs of mega-classes $\nu_t + r + 1, \ldots, \nu_{t+1} - r - 1$.*

3. $W_1^S \leq W_2^S \leq \cdots \leq W_m^S$.

4. *For each pair of jobs $j, j' \notin \cup_t J^t(\bar{a})$ belonging to a common mini-class, if $j < j'$, then $S(j) \leq S(j')$.*

5. *For each pair of alternative jobs $j, j'$ resulting from the set $J^t(\bar{a})$ belonging to a common alternative mini-class such that the size of $j$ is smaller than the size of $j'$, the following holds. If $S$ schedules the original jobs in $j$ and $j'$ on machines $i$ and $i'$, respectively, then $i \leq i'$.*

**Theorem 15** *Given a schedule $S : J \to M$ such that $W_1^S \leq W_2^S \leq \cdots \leq W_m^S$, there exists a structured schedule $S^* : J \to M$ such that for $i = 1, 2, \ldots, m$, we have*

$$(1 - 14\varepsilon) \cdot W_i^S \leq W_i^{S^*} \leq (1 + 14\varepsilon) \cdot W_i^S . \qquad (3)$$

**Proof.** Let $S'$ be the good schedule that is based on $S$ as established in Lemma 12. We apply a sorting procedure of the works of the machines similarly to the one of [15]. In this procedure we are given as an input a partition of the jobs into subsets $\mathcal{J}_1, \ldots, \mathcal{J}_m$, we create a new partition of the jobs as follows. For $i = 1, 2, \ldots, m - 1$ we assume that we are given the subsets $\mathcal{J}_i, \ldots \mathcal{J}_m$ and we choose the set of jobs scheduled on machine $i$ (possibly modifying the remaining subsets). For each mini-class (including the alternative mini-classes), we temporarily replace the jobs in $\mathcal{J}_{i'}$ ($i' = i, i+1, \ldots, m$) from this mini-class with the smallest set of jobs of this mini-class which are still available (i.e., they are not scheduled on machines with indices smaller than $i$). We pick the set of jobs which have minimum total size as the set $\mathcal{J}_i$, possibly swapping locations of jobs in the same mini-class. Note that due to our use of alternative mini-classes, the jobs that are inside these alternative jobs might *not* be allocated in order of their size (but still in a fixed order according to the size of the alternative jobs). Consider a pair of consecutive machines $i, i + 1$, then the resulting work of machine $i$ is not larger than the resulting work of machine $i + 1$, since the set of jobs allocated to machine $i + 1$ were available for allocation to this subset of jobs when we picked $\mathcal{J}_i$ for machine $i$ (the jobs taken by $\mathcal{J}_i$ are replaced by other jobs of the same mini-class when we choose $\mathcal{J}_{i+1}$, which cannot be smaller).

We apply the sorting procedure on the partition defined by $S'$. The output of this procedure is an integral schedule denoted by $S^*$. Clearly, $W_1^{S^*} \leq W_2^{S^*} \leq \cdots \leq W_m^{S^*}$. Moreover, properties 2, 4 and 5 in the definition of structured schedules are satisfied. We next prove (3) for every machine $i$. Every machine $i$ receives a subset of jobs which is based on a subset of jobs allocated to some machine $i'$ in $S'$, after swapping pairs of jobs within a common mini-class. Therefore,

$$\frac{W_{i'}^{S'}}{1 + \varepsilon} \leq W_i^{S^*} \leq (1 + \varepsilon) \cdot W_{i'}^{S'} . \qquad (4)$$

Fix a machine index $i$. When we choose the set $\mathcal{J}_i$ for machine $i$, at least one of the original $i$ subsets $\mathcal{J}_1, \ldots \mathcal{J}_i$ (up to swapping some locations of pairs of jobs within common mini-classes and alternative mini-classes) remains available. Let $i''$ denote its index. The total size of the jobs in this subset is at most $W_{i''}^{S'} \cdot (1+\varepsilon) \leq W_{i''}^{S} \cdot (1+12\varepsilon)(1+\varepsilon) \leq W_i^S \cdot (1+12\varepsilon)(1+\varepsilon)$ where the first inequality holds by Lemma 12, and the second inequality holds by the monotonicity of works in $S$. Therefore, machine $i$ receives in $S^*$ a total work of at most $W_i^S \cdot (1+14\varepsilon)$. We next prove the other inequality, that is, $(1-14\varepsilon) \cdot W_i^S \leq W_i^{S^*}$. The schedule $S$ has at most $i-1$ machines which receive work strictly below $W_i^S$. Therefore, in $S'$ there are at most $i-1$ machines which receive work strictly below $W_i^S \cdot (1-12\varepsilon)$. In $S^*$ the number of machines with work strictly smaller than $\frac{W_i^S \cdot (1-12\varepsilon)}{1+\varepsilon}$ cannot exceed $i-1$, and due to the monotonicity of the works in $S^*$, the claim holds.

Finally, we prove property 1 of structured schedules. The integral schedule $S^*$ is quasi-consistent with the magnitude vector $\bar{a}$. To see this claim, consider a machine $i$ such that $\xi_t \leq i < \xi_{t+1}$ and denote by $j$ the maximum sized job on machine $i$ according to $S^*$. We need to prove that $p_j \leq 2^{\nu_{t+1}+r+1}$. The set of jobs $\mathcal{J}_i$ which the sorting procedure allocated to machine $i$ was scheduled on a machine $i'$ in $S'$ (possibly swapping pairs of jobs in common mini-classes). A job $j'$ of the same mini-class as $j$ was allocated to machine $i'$ in $S'$. Recall that $\frac{W_{i'}^{S'}}{1+\varepsilon} \leq W_i^{S^*} \leq (1+\varepsilon) \cdot W_{i'}^{S'}$. By Lemma 12, $S'$ is almost consistent with $\bar{a}$, and therefore $p_{j'} \leq 2^{a_{i'}+r+1}$. Since $j$ and $j'$ belong to a common mini-class, they also belong to a common mega-class, and thus we also have $p_j \leq 2^{a_{i'}+r+1}$. In order to prove that $p_j \leq 2^{\nu_{t+1}+r+1}$ it suffices to show that $i' < \xi_{t+2}$. Assume by contradiction that $i' \geq \xi_{t+2}$. We have $W_{i'}^S \geq W_{\xi_{t+2}}^S > 2^{\nu_{t+2}-1}$ which holds since the works in $S$ are monotonically non-decreasing and $2W_{\xi_{t+2}}^S > \tilde{W}_{\xi_{t+2}}^S = 2^{\alpha_{\xi_{t+2}}^S} = 2^{\nu_{t+2}}$. On the other hand, $W_i^S \leq W_{\xi_{t+1}}^S \leq 2^{\nu_{t+1}}$. By (4) and Lemma 12, we have $W_i^{S^*} \geq \frac{W_{i'}^{S'}}{1+\varepsilon} \geq \frac{W_{i'}^S(1-12\varepsilon)}{1+\varepsilon} \geq W_{i'}^S(1-14\varepsilon)$. Therefore, we get

$$\begin{aligned}
2^{\nu_{t+1}} &\geq W_i^S \\
&\geq \frac{W_i^{S^*}}{1+14\varepsilon} &&\text{by (3)} \\
&\geq \frac{W_{i'}^S(1-14\varepsilon)}{1+14\varepsilon} \\
&> 2^{\nu_{t+2}-1} \cdot \frac{1-14\varepsilon}{1+14\varepsilon} \\
&> 2^{\nu_{t+2}-3} \geq 2^{\nu_{t+1}+r-2}, &&\text{since } \varepsilon < \frac{1}{32}
\end{aligned}$$

contradicting $r \geq 5$. Therefore, $S^*$ is quasi-consistent with the magnitude vector $\bar{a}$ because for every $i$, we have $|\alpha_i^S - \alpha_i^{S^*}| \leq 1$ since $1+14\varepsilon < 2$ and $\frac{1}{1-14\varepsilon} < 2$.

Fix a value of $t = 1, 2, \ldots, \tau(\bar{a}) - 5$, it remains to prove that there is no $j$ and $i \geq \xi_{t+5}$ such that $p_j \leq 2^{\nu_t - r}$ and $S^*$ schedules job $j$ to machine $i$.

Consider a machine $i$ such that $\xi_{t+5} \leq i < \xi_{t+6}$. The set of jobs $\mathcal{J}_i$ which the sorting procedure allocated to machine $i$, was scheduled on a machine $i'$ in $S'$ (possibly swapping pairs of jobs in common mini-classes). Let $j$ be a job such that $S^*(j) = i$. In order to prove that $p_j > 2^{\nu_t - r}$ it suffices to show that $i' \geq \xi_{t+4}$. Assume by contradiction that $i' < \xi_{t+4}$. We have $W_{i'}^S \leq W_{\xi_{t+4}}^S \leq 2^{\nu_{t+4}}$ which holds since the works in $S$ are monotonically non-decreasing and $W_{\xi_{t+4}}^S \leq \tilde{W}_{\xi_{t+4}}^S = 2^{\alpha_{\xi_{t+4}}^S} = 2^{\nu_{t+4}}$. On the other hand, $W_i^S \geq W_{\xi_{t+5}}^S > 2^{\nu_{t+5}-1}$. By (4) and Lemma 12, we have $W_i^{S^*} \leq W_{i'}^{S'} \cdot (1+\varepsilon) \leq W_{i'}^S(1+12\varepsilon)(1+\varepsilon) \leq$

$W_{i'}^S(1+14\varepsilon)$. Therefore, using (3) we get

$$
\begin{aligned}
2^{\nu_{t+5}} &\leq 2W_i^S \\
&\leq \frac{2W_i^{S^*}}{1-14\varepsilon} && \text{by (3)} \\
&\leq \frac{2W_{i'}^S(1+14\varepsilon)}{1-14\varepsilon} \\
&\leq 2^{\nu_{t+4}+1} \cdot \frac{1+14\varepsilon}{1-14\varepsilon} \\
&< 2^{\nu_{t+4}+3} \leq 2^{\nu_{t+5}-r+2}, && \text{since } \varepsilon < \frac{1}{32}
\end{aligned}
$$

contradicting $r \geq 5$. ∎

## 4 A dynamic programming for computing the best highly structured solution

In this section we show how to compute the optimal structured schedule $S^*$. Our algorithm will use a dynamic programming procedure which is based on a shortest path (or an optimal bottleneck path) in a directed layered graph $G = (V, E)$ with weights on its vertices.

We will define a layered graph, in which the algorithm computes a path corresponding to an optimal solution with respect to a given goal function. Each layer of an index $1, 2, \ldots, m$ corresponds to a machine, and each vertex in one of these layers encodes a set of jobs which were scheduled prior to the current machine, and a set of jobs which were scheduled up to and including the current machine. The difference between these sets easily reveals the work of the current machine, and allows us to restrict the paths in the graph to schedules in which the works are monotonically non-decreasing. Given the work $W_i$ of the current machine $i$, the weight of the vertex is the load of this machine $L_i$, or $f(L_i)$ for a well-behaved function $f$. The edges between layers correspond to compatibility conditions which in particular enforce the condition that the works of machines are monotonically non-decreasing. The order of the layers is according to the speeds of the machines, that is, machines with higher speeds have a higher index of their layers, and subsets of machines with a common speed are ordered according to a fixed ordering of the machines. The graph which we will use allows us to find any structured schedule and maybe additional schedules. The schedule which will be found will be at least as good as the structured schedule whose existence we proved in the previous section. To distinguish between several optimal solutions, and to prioritize the possible outputs, we number all vertices of each layer with distinct integers, and we always search for paths whose reverse sequence of numbers along the path (that is, the sequence of vertices given from the end of the path towards the beginning of the path) is minimal (lexicographically) out of paths which have an optimal cost with respect to our goal function. This property allows us to assume that there exists a total order over the paths in the graph, and the algorithm always outputs the minimal path (according to this order) which is optimal in the current scenario.

The graph $G$ will encode in each layer all possible short histories of the magnitude vectors (which we call short magnitude vectors). There will be a starting vertex $s$, also seen as the layer of vertices of index 0, and an end vertex $t$, also seen as the layer of vertices of index $m+1$, and we always look for a path in $G$ from $s$ to $t$. Thus, $V$ consists of $m$ regular layers denoted as $1, 2, \ldots, m$ (one for each machine) and two additional layers $0$ and $m + 1$. For every possible structured schedule, there will be an $s - t$ path corresponding to it (and possibly additional $s - t$ paths corresponding to other feasible schedules).

A *short magnitude vector* $\psi = (\psi^0, \psi^1, \psi^2, \ldots, \psi^6)$ for machine $i$ is a vector consisting of seven consecutive distinct values in a magnitude vector $\bar{a}$ (that is, there exists $1 \leq t \leq \tau(\bar{a})$ such that $\psi^\eta = \nu_{t+\eta-5}$ for $\eta = 0, 1, 2, \ldots, 6$). If this vector is associated with machine $i$, then $a_i = \psi^5$. If $a_i = \nu_{t+5}$ for some value of $t$, then $\psi = (\nu_t, \nu_{t+1}, \ldots, \nu_{t+6})$. If the magnitude $\psi^5$ is the largest magnitude in $\bar{a}$, we will let $\psi^6$ be the fictitious value $\infty$. Similarly, if $\psi^5$ is one of the smallest five values in $\bar{a}$, we add $-\infty$ as the first components of $\psi$. We say that a short magnitude vector $\psi$ is quasi-consistent with a schedule $S$ is $\psi$ consists of six consecutive distinct values of a magnitude vector $\bar{a}$ such that $S$ is quasi-consistent with $\bar{a}$.

Other than the entries which are $-\infty$ or $\infty$, a short magnitude vector must be such that it can be a part of a magnitude vector. Thus, $\psi^{\eta+1} \geq \psi^\eta + r + 1$ for $\eta = 0, \ldots, 5$. In addition, we define a list of allowed finite components.

**Lemma 16** *For $j \in J$ let $\tilde{p}_j = \lceil \log_2 p_j \rceil$. Then, for every possible subset $J' \subseteq J$ of jobs whose total size is $W$, we have*

$$\lceil \log_2 W \rceil \in \bigcup_{j \in J} \bigcup_{k=0}^{\lceil \log_2 n \rceil} \{\tilde{p}_j + k\} .$$

**Proof.** Let $j \in J'$ be a maximum indexed job in $J'$. Then, $W \geq p_j$ and $W \leq j \cdot p_j \leq n \cdot p_j$. Therefore, $\tilde{p}_j \leq \lceil \log_2 W \rceil \leq \lceil \log_2(n \cdot p_j) \rceil = \lceil \log_2 n + \log_2 p_j \rceil \leq \lceil \log_2 n \rceil + \tilde{p}_j$ and the claim holds. ∎

**Corollary 17** *The number of possibilities of short magnitude vectors is $O(n^8)$.*

**Proof.** The set of different values for each component in a short magnitude vector is

$$\bigcup_{j \in J} \bigcup_{k=-1}^{\lceil \log_2 n \rceil + 1} \{\tilde{p}_j + k\} \cup \{-\infty, \infty\}$$

since we are only interested in magnitude vectors which are quasi-consistent with some schedule. Therefore, there are at most $(n \cdot (\log_2 n + 4) + 2)^7 = O(n^8)$ different short magnitude vectors. ∎

Next, we define the set $A(\psi)$ of active mega-classes for a short magnitude vector $\psi$. A mega-class $k$ belongs to $A(\psi)$ if there exists a value of $\eta = 0, 1, \ldots, 6$ such that $|k - \psi^\eta| \leq r$, and an alternative mega-class $\psi^{\eta+1} - r - 1$ (and perhaps a smaller alternative mega-class consisting of a single alternative job) belongs to $A(\psi)$ if it is an alternative mega-class consisting of alternative jobs of mega-classes $\psi^\eta + r + 1, \ldots, \psi^{\eta+1} - r - 1$ for values of $\eta = 0, 1, \ldots, 5$ for which $\psi^{\eta+1} - \psi^\eta \geq 2r + 2$.

The motivation for this definition of $A(\psi)$ is that if machine $i$ has a short magnitude vector $\psi$, then all jobs of size at most $2^{\psi_0 - r}$ are scheduled on machines with magnitude at most $\psi^4$ in any structured schedule that is quasi-consistent with $\psi$, i.e., strictly before machine $i$ (since $a_i = \psi^5$). Moreover, all jobs of size more than $2^{\psi_6 + r + 1}$ are scheduled on machines with magnitude at least $\psi^6$ in any structured schedule that is quasi-consistent with $\psi$, i.e., after machine $i$. Thus the only relevant (alternative) mega-classes for machine $i$ are the ones described above. These properties will be enforced by the structure of the graph. Moreover, given a set of consecutive mega-classes it can be decided to convert the jobs of these mega-classes into alternative jobs, and this can only happen if no jobs of these mega-classes were already scheduled. Once it is decided, this decision is irrevocable and future sets of consecutive mega-classes which are converted into alternative jobs will be disjoint.

A *status vector* of a short magnitude vector $\psi$ consists of a component for each mini-class which belongs to a mega-class in $A(\psi)$. This component represents the number of jobs (or alternative jobs if this is an alternative mini-class) which were already scheduled (recall that in a structured schedule we always schedule these jobs sorted by their sizes (with a fixed tie-breaking policy), and therefore the number of jobs which were scheduled uniquely identifies which jobs these are).

**Lemma 18** *The number of status vectors for one specific short magnitude vector $\psi$ is $O(n^{(7(2r+1)+12)\lambda})$. Therefore, overall there are $O(n^{((14r+19)\lambda+8)})$ status vectors.*

**Proof.** The claim holds since every component in the status vector is an integer in $[0, n]$, the number of mini-classes in a mega-class is $\lceil \log_{1+\varepsilon} 2 \rceil = \lambda$, and there are at most 12 alternative mega-classes in $A(\psi)$. ∎

Consider a pair of ordered pairs $(\psi, u)$ and $(\psi', u')$ where $u$ and $u'$ are status vectors of the short magnitude vectors $\psi$ and $\psi'$, respectively. We say that such a pair is compatible if one of the following cases hold.

1. If $\psi = \psi'$ and every component in $u$ is at most its corresponding component in $u'$.

2. If for all $\eta = 1, 2, \ldots, 6$, $\psi^\eta = (\psi')^{\eta-1}$, and every component in $u$ corresponding to a mini-class $(k, \ell)$ (such that mega-class $k$ is in $A(\psi')$) is at most its corresponding component in $u'$. Moreover, every component in $u'$ which corresponds to a mini-class $(k, \ell)$ such that $k \notin A(\psi)$ is zero. Informally, jobs of such zero components in $u'$ are too large for $\psi$.

If $(\psi, u)$ and $(\psi', u')$ are compatible, then their difference defines a set of jobs which can be scheduled on a machine. This set of jobs $J((\psi, u), (\psi', u'))$ is defined as follows. The set $J((\psi, u), (\psi', u'))$ will contain all remaining jobs of mini-classes which have corresponding components in $u$ but not in $u'$ (these are the last jobs of each mini-class which are not scheduled yet, according to the information encoded in $u$). Informally, such jobs are too small for $\psi'$ and must be assigned immediately. For every mini-class which has components in both $u$ and $u'$, the number of jobs of this mini-class in $J((\psi, u), (\psi', u'))$ is the difference between these components (these are the next jobs in each mini-class). We denote by $W((\psi, u), (\psi', u'))$ the total size of jobs in $J((\psi, u), (\psi', u'))$.

The set of vertices of layer $i$ (for $i = 1, 2, \ldots, m$) is the set of compatible pairs $(\psi, u)$ and $(\psi', u')$. Thus such a vertex corresponds to $((\psi, u), (\psi', u'))$. The meaning of such a pair is to assign the jobs of their difference to machine $i$ (and thus the work of $i$ would be exactly $W((\psi, u), (\psi', u'))$), where $\psi$ is the short magnitude vector of machine $i$, and $\psi'$ is the short magnitude vector of machine $i + 1$.

The weight of such a vertex in layer $i$ is defined as $\frac{W((\psi,u),(\psi',u'))}{s_i}$ if we are solving the minimum makespan problem or the problem of maximizing the minimum load. If we are interested in the problem of minimizing $\sum_{i=1}^m f(L_i)$ for a well-behaved function $f$, then the weight of the vertex is $f(\frac{W((\psi,u),(\psi',u'))}{s_i})$. The vertices $s, t$ do not have weights.

A vertex in layer $i$ (for $1 \leq i \leq m-1$) corresponding to $((\psi, u), (\psi', u'))$ is adjacent to a vertex in layer $i+1$ corresponding to $((\psi', u'), (\psi'', u''))$ if and only if $W((\psi, u), (\psi', u')) \leq W((\psi', u'), (\psi'', u''))$. There are no other edges between these layers, that is, there can be no edge from $((\psi^1, u^1), (\psi'^1, u'^1))$ to $((\psi^2, u^2), (\psi'^2, u'^2))$ in consecutive layers if $(\psi'^1, u'^1) \neq (\psi^2, u^2)$. The vertex $s$ of layer 0 is adjacent to all vertices in layer 1 corresponding to $((\psi, u), (\psi', u'))$ such that all components of the status vector $u$ are zero, and $\psi^0 = -\infty$. The vertices of layer $m$ which are adjacent to $t$ (of layer $m+1$) are the ones corresponding to $((\psi, u), (\psi', u'))$ such that $\psi'^6 = \infty$, and for every mini-class whose mega-class is in $A(\psi')$ the component in $u'$ is exactly the number of jobs in this mini-class (also for an alternative mini-class). The topology of the graph $G$ depends only on the set of jobs and their sizes, and on the number of machines (and not on their speeds). Only the weights depend on the exact problem and on the speeds of the machines.

We observe that an $s-t$ path in the graph gives immediately a schedule, since each vertex $((\psi, u), (\psi', u'))$ in the graph defines a specific set of jobs allocated to the machine with index equal to the index of its layer, whose total size is exactly $W((\psi, u), (\psi', u'))$. Moreover, every $s-t$ path defines a partition of the job set, and every such solution satisfies that the works of the machines are monotonically non-decreasing in the index of the machine. We also observe that every structured solution corresponds to (at least) one $s-t$ path in the graph $G$.

Using this graph, we compute a label for each vertex. This label is equal to the cost (or value) of the partial solution defined by the best path from $s$ to this vertex. Moreover, we compute a pointer $\pi$ to the previous vertex on this best path from $s$. If there are several possibilities for best paths (ending at the same vertex) $\pi$ is defined to be the minimum index of the vertex satisfying these conditions according to the numbering of vertices in each layer. We next define the notion of a best path for each of the objectives considered in this paper. For the problem of minimizing the makespan, a best path is one that minimizes the maximum weight of a vertex along the path. For the problem of maximizing the minimum load, a best path is one that maximizes the minimum weight of a vertex along the path. Finally for the problem of minimizing $\sum_{i=1}^{m} f(L_i)$ where $f$ is a well-behaved function, a best path is a path of minimum total weight of its vertices.

## 5 Monotonicity proof

Our monotonicity proofs are based on analysis of a scenario where machine $\gamma$ changes its speed. We will assume that every machine $\gamma' \neq \gamma$ has a fixed speed of $s_{\gamma'}$ while machine $\gamma$ has two possible speeds $s_\gamma$ and $s'_\gamma$. We sometimes consider additional speeds between $s_\gamma$ and $s'_\gamma$. In the next two lemmas $s_1, \ldots, s_m$ denotes a sorted list of machines speeds.

**Lemma 19** *Consider two executions of the algorithm, both with respect to minimizing $\sum_{i=1}^{m} f(L_i)$ where $f$ is a well-behaved function (with a common function $f$), where the sorted order of machines is $1, 2, \ldots, m$, each with its own set of speeds, resulting in the two schedules $S_1$ and $S_2$ found by the paths $P_1$ and $P_2$. The two sets of speeds are defined as follows. For every $i' \neq i$ the speed of $i'$ is $s_{i'}$ in both sets, and the speed of $i$ is $\sigma_1$ and $\sigma_2$, respectively, such that $s_{i-1} \leq \sigma_1 < \sigma_2 \leq s_{i+1}$ (where $s_0 = 0$ and $s_{m+1} = \infty$). Then, $W_i^{S_1} \leq W_i^{S_2}$.*

**Proof.** For a schedule $S$ denote by $\text{COST}_S, \text{COST}'_S$ the costs of schedule $S$ using the speeds $\sigma_1$ and $\sigma_2$ for machine $i$, respectively. Recall that the graph $G$ remains the same in the two executions. Since the path $P_1$ could have been found by the algorithm when it computes $P_2$ and vice versa, $\text{COST}'_{S_1} \geq \text{COST}'_{S_2}$, and $\text{COST}_{S_2} \geq \text{COST}_{S_1}$, which gives $\text{COST}'_{S_1} - \text{COST}_{S_1} \geq \text{COST}'_{S_2} - \text{COST}_{S_2}$. Assume by contradiction $W_i^{S_2} < W_i^{S_1}$.

Since $\sigma_2 > \sigma_1$ and $W_i^{S_2} < W_i^{S_1}$, we find $W_i^{S_1}(\frac{1}{\sigma_1} - \frac{1}{\sigma_2}) > W_i^{S_2}(\frac{1}{\sigma_1} - \frac{1}{\sigma_2})$. Rearranging the last inequality gives $\frac{W_i^{S_1}}{\sigma_1} + \frac{W_i^{S_2}}{\sigma_2} - \frac{W_i^{S_1}}{\sigma_2} > \frac{W_i^{S_2}}{\sigma_1}$. Since $i$ does not change its position in the sorted order of machines, we have $\text{COST}'_{S_1} - \text{COST}_{S_1} = f(\frac{W_i^{S_1}}{\sigma_2}) - f(\frac{W_i^{S_1}}{\sigma_1})$ and $\text{COST}'_{S_2} - \text{COST}_{S_2} = f(\frac{W_i^{S_2}}{\sigma_2}) - f(\frac{W_i^{S_2}}{\sigma_1})$, and so we find using $\text{COST}'_{S_1} - \text{COST}_{S_1} \geq \text{COST}'_{S_2} - \text{COST}_{S_2}$ that $f(\frac{W_i^{S_1}}{\sigma_2}) - f(\frac{W_i^{S_1}}{\sigma_1}) \geq f(\frac{W_i^{S_2}}{\sigma_2}) - f(\frac{W_i^{S_2}}{\sigma_1})$. Using $\sigma_1 < \sigma_2$ and $W_i^{S_2} < W_i^{S_1}$ we have $\frac{W_i^{S_2}}{\sigma_2} < \frac{W_i^{S_1}}{\sigma_2} < \frac{W_i^{S_1}}{\sigma_1}$. By convexity, we find $f(\frac{W_i^{S_1}}{\sigma_1}) + f(\frac{W_i^{S_2}}{\sigma_2}) \geq f(\frac{W_i^{S_1}}{\sigma_2}) + f(\frac{W_i^{S_1}}{\sigma_1} + \frac{W_i^{S_2}}{\sigma_2} - \frac{W_i^{S_1}}{\sigma_2})$. Using strict monotonicity of $f$, $f(\frac{W_i^{S_1}}{\sigma_1} + \frac{W_i^{S_2}}{\sigma_2} - \frac{W_i^{S_1}}{\sigma_2}) > f(\frac{W_i^{S_2}}{\sigma_1})$, which is a contradiction. ∎

The next lemma is used in the case that the speed of a machine changes. We will split the process of changing the speed into steps, and one type of step will be swapping the positions with another machine of the same speed. Therefore, we note the following.

**Lemma 20** *Consider two executions of the algorithm, both with respect to the same objective function, each with the same set of speeds $s_1, s_2, \ldots, s_m$ where $s_i = s_{i+1}$, where the sorted order of machines is given by increasing indices in the first execution and the order obtained by swapping the positions of machines $i, i+1$ in the second execution, resulting in the two schedules $S_1$ and $S_2$. Denote by $\omega_1$ the work of machine $i$ in the schedule $S_1$ (that is, $\omega_1 = W_i^{S_1}$), and by $\omega_2$ the work of the same machine in $S_2$ (that is, $\omega_2 = W_{i+1}^{S_2}$). Then, $\omega_1 \leq \omega_2$.*

**Proof.** Since the set of optimal solutions for the two inputs is exactly the same, so is the set of optimal paths in the graph. Since our algorithm always outputs lexicographic minimal optimal path, we conclude that $S_1 = S_2$. The claim holds because the solutions obtained as paths in the graph have monotonically non-decreasing works of machines. ∎

**Theorem 21** *The approximation scheme for minimizing $\sum_{i=1}^m f(L_i)$ where $f$ is a well-behaved function is a monotone PTAS. The approximation scheme for minimizing the $\ell_p$ norm of the vector of machine loads (obtained by running the algorithm with $f(x) = x^p$) is a monotone PTAS even if $p$ is a part of the input.*

**Proof.** Let $S$ be an optimal solution, then by Theorem 15, there is a structured solution $S^*$ such that for every $i$ we have $W_i^{S^*} \leq (1+14\varepsilon) \cdot W_i^S$, and thus $L_i^{S^*} \leq L_i^S \cdot (1+14\varepsilon)$, and therefore the cost of $S^*$ as a solution to our problem is at most $\sum_{i=1}^m f(L_i^{S^*}) \leq \sum_{i=1}^m f(L_i^S \cdot (1+14\varepsilon)) \leq (1+O(1)\varepsilon) \sum_{i=1}^m f(L_i^S)$ where the first inequality holds by monotonicity of $f$, and the second inequality by the property of $f$ that if $x \leq (1+\varepsilon)y$ then $f(x) \leq (1+O(1)\varepsilon)f(y)$. The schedule given by the algorithm as output has a cost which is no larger than the cost of $S^*$. Note that the approximation ratio of $S^*$ for the problem of minimizing the $\ell_p$ norm of the vector of machine loads is at most $1 + 14\varepsilon$ since $\left(\sum_{i=1}^m \left(L_i^{S^*}\right)^p\right)^{1/p} \leq (1+14\varepsilon) \cdot \left(\sum_{i=1}^m \left(L_i^S\right)^p\right)^{1/p}$.

To prove the monotonicity, consider a machine $i$ which increases its speed from $s_i$ to $s_i'$. We split the process of increasing the speed of a given machine into two types of events. The first type are time intervals in which the position of this machine in the sorted order of the machines does not change. The second type are points in time when the speed is fixed, but the machine swaps its location with the next machine in the list of machines sorted by speed. There can be multiple such time intervals and points in time, and it is sufficient to consider one event of each type, thus we consider two cases. The case where machine $i$ increases its speed, $s_i' \leq s_{i+1}$, and machine $i$ does not change its position in the sorted list of machines, and the case $s_i = s_{i+1}$, where the only change is that these two machines swap their relative order. For the first case, the claim follows by Lemma 19. For the second case, the claim follows by Lemma 20. ∎

**Theorem 22** *The approximation scheme for maximizing $\min_{i \in M} L_i$ is a monotone PTAS.*

**Proof.** Let $S$ be an optimal solution, then by Theorem 15, there is a structured solution $S^*$ such that for every $i$ we have $W_i^{S^*} \geq (1 - 14\varepsilon) \cdot W_i^S$, and thus $L_i^{S^*} \geq L_i^S \cdot (1 - 14\varepsilon)$, and therefore the value of $S^*$ is at least $1 - 14\varepsilon$ times the value of $S$. The schedule given by the algorithm as output has a value which is no smaller than the value of $S^*$.

To prove the monotonicity, consider a machine $i$ which increases its speed from $s_i$ to $s_i'$. Consider the solution $S_1$ obtained by the algorithm for the case where the speed of $i$ is $s_i$. Let $C_1$ be the value of $S_1$ (computed for the set of speeds where the speed of $i$ is $s_i$). We split the process of increasing the speed of a given machine into two periods where the first period is split further into two types of events. In the first period, the speed of $i$ is at most $\sigma$, where $\sigma$ is the maximum speed for which the value of the solution $S_1$ is exactly $C_1$ (possibly swapping the contents of machines if machine $i$ changes its position in the sorted list of machines according to the sorting done by the algorithm). Note that $\sigma$ is well-defined, that is, the maximum exists. If $\sigma = s_i$, we say that this period is empty. If $\sigma > s_i'$, we set $\sigma = s_i'$. Therefore, during the first period the speed of $i$ is in $(s_i, \sigma]$. If $\sigma = s_i'$, then the second period is empty, and otherwise the speed of $i$ is in $(\sigma, s_i']$ in this period. For the first period, the first type of events are time intervals in which the position of this machine in the sorted order of the machines does not change. The second type are points in time when the speed is fixed, but the machine swaps its location with the next machine in the list of machines sorted by speed.

We prove that for every speed in $[s_i, \sigma]$, the solution $S_1$ is returned by the algorithm. First, we show that the value of an optimal solution remains $C_1$. The value of an optimal solution cannot increase when $i$ increases its speed, so by definition $S_1$ remains an optimal solution. Moreover, when $i$ increases its speed in the first period, the set of optimal solutions is a subset of the set of optimal solutions when the speed of $i$ is $s_i$ (even if locations of machines are swapped). Therefore, the algorithm outputs $S_1$ for every speed in the first period. Thus, for time intervals in which the position of $i$ in the sorted list of machines is fixed, the work of $i$ is exactly the same, and in events in which machine $i$ swaps its position with another machine, the work of $i$ cannot decrease by Lemma 20. In the case $\sigma = s_i'$ we are done. Otherwise, we assume that there are no further machines of speed $\sigma$ which appear later than $i$ in the ordering of the machines (possibly by adding events of the second type for the first period).

Next consider the case where $\sigma < s_i'$. Denote by $W$ the work of $i$ in the solution $S_1$ where the speed of $i$ is $\sigma$. Recall that for this speed of $i$, the value of the optimal solution (i.e., of $S_1$) is exactly $C_1$. We prove that $\frac{W}{\sigma} = C_1$. Assume by contradiction that the claim does not hold (that is we assume that $\frac{W}{\sigma} > C_1$, as otherwise the value of $S_1$ in this case is strictly smaller than $C_1$ contradicting the definition of $\sigma$). Let $\sigma_1 > \sigma$ be such that $\sigma_1 \leq \frac{W}{C_1}$ and $\sigma_1$ is smaller than the speed of the next machine after $i$ in the sorted list of machines, if such a machine exists. Then, the value of $S_1$ for the speed $\sigma_1$ of $i$ remains $C_1$ contradicting the maximality of $\sigma$. Let $C_2$ be the value of an optimal solution $S_2$ found by the algorithm where the speed of $i$ is $s_i'$. Then, $C_2 \geq C_1 \cdot \frac{\sigma}{s_i'}$ since otherwise $S_1$ is a strictly better solution for speed $s_i'$ of $i$, because even if machines swap locations the machine in every position is faster by no more than $\frac{s_i'}{\sigma}$. Denote by $W'$ the work of $i$ in $S_2$. We have $W' \geq C_2 \cdot s_i' \geq C_1 \cdot \sigma = W$, and the claim follows. ∎

The proof of the next theorem is similar to the proof of Theorem 22, and it is given in Appendix A.2.

**Theorem 23** *The approximation scheme for minimizing $\max_{i \in M} L_i$ is a monotone PTAS.*

## 6 Computing the payments

Archer and Tardos [5] defined a payment scheme which can be applied for any monotone scheduling algorithm to create a truthful mechanism. Denote the payment to agent $i$ by $P_i$. We briefly repeat the definition of $P_i$. Let $b_{-i}$ denote the vector of bids, not including agent $i$. We write $b$ (the complete bid vector) also as $(b_{-i}, b_i)$. Then the payment function for agent $i$ is defined as

$$P_i(b_{-i}, b_i) = h_i(b_{-i}) + b_i w_i(b_{-i}, b_i) - \int_0^{b_i} w_i(b_{-i}, u) du, \tag{5}$$

where $w_i(b_{-i}, b_i)$ is the work (total size of jobs) allocated to machine $i$ given the bid vector $b$ and the $h_i$ are arbitrary functions (Theorem 4.2 in [5]).

In order to compute the payments, we need to calculate the integral in (5). Recall that the bid of an agent represents its claimed cost for processing one unit of work, which can be seen as the inverse of the speed of its machine. For a given set of bids $(b_1, \ldots, b_m)$, calculating the integral for agent $i$ requires us to know what its work would be for every possible bid $b$ of this agent, i.e., for the bids $(b_{-i}, b)$ for $b \in (0, \infty)$. First, we partition the possible bids into intervals in which the position in the ordered set of machines (that is, its layer in the graph) of machine $i$ remains constant. Consider the set $\{0, \infty\} \cup \{b_j\}_{j=1}^m \setminus \{b_i\}$ and denote its elements by $0 = c_1 < \cdots < c_{m'} = \infty$ ($m' \leq m+1$), then the intervals to consider are $(c_j, c_{j+1})$ for $j = 1, \ldots, m'-1$.

For each vertex $v$ in layer $i$, we compute a function $F_v(b)$ which is the objective function value of the best path which traverses *this vertex*, as a function of the bid of machine $i$. Recall that the algorithm outputs the minimum or maximum (over all vertices of the layer) of these functions depending on the objective function.

**Claim 24** *For each vertex $v$ and every bid interval $(c_j, c_{j+1})$, $F_v(b)$ is a piecewise linear continuous function with a polynomial number of pieces.*

**Proof.** In layer $i$, the weight of vertex $v$ which represents the compatible pair $((\psi, u), (\psi', u'))$ is the constant $W((\psi, u), (\psi', u'))$ divided by $s_i$, where $s_i = 1/b_i$. Note that the pair represented by $v$ also specifies the set of jobs assigned to machines before machine $i$, and the set assigned after $i$. Due to the tie breaking done in the dynamic program, and the fact that only the speed of machine $i$ changes, this means that the identity of the best path which passes through $v$ does not depend on $b$ (only its objective value does).

For the makespan and the maximizing the minimum load problems, the objective value of a path is the maximum (minimum, respectively) weight of a vertex along the path. Hence, as $b_i$ increases from $c_j$ to $c_{j+1}$, the only change that can happen is that the weight of vertex $v$ starts having the maximum weight along the fixed best path (for the makespan objective) or stops having the minimum weight (for the covering objective). Therefore, $F_v(b)$ has at most two pieces, where for one piece machine $i$ is a *bottleneck* machine (that is, a machine whose load equals the objective function value of the solution) and for the other it is not. If $i$ is the bottleneck, $F_v(b) = b \cdot W((\psi, u), (\psi', u'))$, else $F_v(b)$ is constant.

For the minimization of $\sum_{i=1}^m f(L_i)$ for a well-behaved function $f$, the objective value of a path is the total weight of its vertices. Here, $F_v(b)$ is a constant plus $f(b \cdot W((\psi, u), (\psi', u')))$ (where the constant is the total weight of the other vertices along the best path which traverses $v$). Hence by using the approximated piecewise-linear convex monotonically increasing function of $f$ instead of $f$ itself the claim follows since it is sufficient to consider such an approximated function with pieces ending at integer powers of $(1+\varepsilon)$ (and thus with polynomially many such pieces). ∎

Claim 24 implies that the number of intersection points between any pair of functions $(F_v(b), F_u(b))$ is also polynomial. Thus we can compute all of these points in polynomial time, and determine which points lie inside the interval $(c_j, c_{j+1})$. Moreover, we can also determine which *schedule* our mechanism uses for each intersection point by running the PTAS for each point, including $c_j$ (if $c_j > 0$) and $c_{j+1}$ (if $c_{j+1} < \infty$). After removing duplicates, this gives us a list of intersection points with associated schedules and works.

**Remark 25** *The replacement of $f$ with the convex monotonically increasing piecewise-linear approximation of $f$ is crucial. Without it, computing the value of $b_i$ in which one solution becomes better than another solution involves computation of an exact solution of equations involving convex functions (this cannot be done even for the case where $f(x) = x^5$). However, for piecewise-linear functions this can be done efficiently.*

It is now straightforward to determine the schedule used for any possible bid $b$, and from that the work for any bid, as follows. Note that the schedule chosen does not change between any pair of consecutive intersection points by construction. Thus the work remains constant between any such pair. If the schedule used is the same at both endpoints, the work in between is given by this schedule. If two different schedules are used, then in the entire open interval between the pair, the schedule is used which gives the best value for the objective function. This can be determined by running the PTAS for one point inside this interval. Thus we can find the exact value of the integral in (5) (without rounding the speeds of the machines).

## A  Omitted proofs

### A.1  Proof of the second part of Claim 1

Consider a schedule $S$ with makespan $M$ and cover $C$. Call a pair of machines $i, j$ *reversed* if $1 \leq i < j \leq m$ and $W_i^S > W_j^S$. We show that removing a consecutive reversed pair (that is, $j = i + 1$) by swapping the sets of jobs assigned to them from any schedule $S$ does not increase the makespan or decrease the cover, which implies the claim (since after a finite number of such steps there will no longer be reversed pairs). Let $S'$ be the schedule resulting from swapping the two job sets of machines $i, j$. In $S'$, machine $j$ gets more work, but the load remains at most $M$: we have $W_j^{S'}/s_j = W_i^S/s_j \leq W_i^S/s_i \leq M$. Machine $i$ gets less work, but the cover remains at least $C$: we have $W_i^{S'}/s_i = W_j^S/s_i \geq W_j^S/s_j \geq C$.

### A.2  Proof of Theorem 23

Let $S$ be an optimal solution, then by Theorem 15, there is a structured solution $S^*$ such that for every $i$ we have $W_i^{S^*} \leq (1 + 14\varepsilon) \cdot W_i^S$, and thus $L_i^{S^*} \leq L_i^S \cdot (1 + 14\varepsilon)$, and therefore the makespan of $S^*$ is at most $1 + 14\varepsilon$ times the makespan of $S$. The schedule given by the algorithm as output has a makespan which is no larger than the makespan of $S^*$.

To prove the monotonicity, consider a machine $i$ which decreases its speed from $s_i$ to $s_i'$. Consider the solution $S_1$ obtained by the algorithm for the case where the speed of $i$ is $s_i$. Let $C_1$ be the makespan of $S_1$ (computed for the set of speeds where the speed of $i$ is $s_i$). We split the process of decreasing the speed of a given machine into two periods where the first period is split further into two types of events. In the first period, the speed of $i$ is at least $\sigma$, where $\sigma$ is the minimum speed for which the makespan of the solution $S_1$ is exactly $C_1$ (possibly swapping the contents of machines if machine $i$ changes its position in the sorted list of machines according to the sorting done by the algorithm). Note that $\sigma$ is well-defined, that is, the minimum exists. If $\sigma = s_i$, we say that this period is empty. If $\sigma \leq s'_i$, we set $\sigma = s'_i$. Therefore, during the first period the speed of $i$ is in $[\sigma, s_i)$. If $\sigma = s'_i$, the second period is empty, otherwise the speed of $i$ is in $[s'_i, \sigma)$. For the first period, the first type of events are time intervals in which the position of this machine in the sorted order of the machines does not change. The second type are points in time when the speed is fixed, but the machine swaps its location with the previous machine in the list of machines sorted by speed.

We prove that for every speed in $[\sigma, s_i]$, the solution $S_1$ is returned by the algorithm. First, we show that the makespan of an optimal solution remains $C_1$. The makespan of an optimal solution cannot decrease when $i$ decreases its speed, and by definition $S_1$ remains an optimal solution. Moreover, when $i$ decreases its speed in the first period, the set of optimal solutions is a subset of the set of optimal solutions when the speed of $i$ is $s_i$ (even if locations of machines are swapped). Therefore, the algorithm outputs $S_1$ for every speed in the first period. Thus, for time intervals in which the position of $i$ in the sorted list of machines is fixed, the work of $i$ is exactly the same, and in events in which machine $i$ swaps its position with the previous machine, the work of $i$ cannot increase by Lemma 20. In the case $\sigma = s'_i$ we are done. Otherwise, we assume that there are no further machines of speed $\sigma$ which appear earlier than $i$ in the ordering of the machines (possibly by adding events of the second type for the first period).

Next consider the case where $\sigma > s'_i$. Denote by $W$ the work of $i$ in the solution $S_1$ where the speed of $i$ is $\sigma$. Recall that for this speed of $i$, the makespan of the optimal solution (i.e., of $S_1$) is exactly $C_1$. We prove that $\frac{W}{\sigma} = C_1$. Assume by contradiction that the claim does not hold (that is we assume that $\frac{W}{\sigma} < C_1$, as otherwise the makespan of $S_1$ in this case is strictly larger than $C_1$ contradicting the definition of $\sigma$). Let $\sigma_1 < \sigma$ be such that $\sigma_1 \geq \frac{W}{C_1}$ and $\sigma_1$ is larger than the speed of the previous machine before $i$ in the sorted list of machines, if such a machine exists. Then, the makespan of $S_1$ for the speed $\sigma_1$ of $i$ remains $C_1$ contradicting the minimality of $\sigma$. Let $C_2$ be the makespan of an optimal solution $S_2$ found by the algorithm where the speed of $i$ is $s'_i$. Then, $C_2 \leq C_1 \cdot \frac{\sigma}{s'_i}$ since otherwise $S_1$ is a strictly better solution for speed $s'_i$ of $i$, because even if machines swap locations the machine in every position is slower by no more than $\frac{\sigma}{s'_i}$. Denote by $W'$ the work of $i$ in $S_2$. We have $W' \leq C_2 \cdot s'_i \leq C_1 \cdot \sigma = W$, and the claim follows.